\documentclass[a4paper,11pt]{article}
\pdfoutput=1 
\usepackage{jheppub} 

\usepackage{graphicx} 
\usepackage{amsmath,amsfonts}  
\usepackage{hyperref} 
\usepackage{cleveref}
\usepackage{mathrsfs}
\usepackage{subfigure}
\usepackage{comment}
\usepackage[normalem]{ulem}
\usepackage{placeins}
\usepackage{amsmath,amssymb,amsfonts,bbold}
\usepackage{braket}
\usepackage{natbib}
\usepackage[dvipsnames,svgnames,x11names]{xcolor}
\usepackage{graphicx,subfigure}
\usepackage{tikz}
\usetikzlibrary{patterns,decorations.pathmorphing,arrows.meta,positioning}
\usepackage{mathtools}

\newcommand{\be}{\begin{equation}}
\newcommand{\ee}{\end{equation}}
\newcommand{\bse}{\begin{subequations}}
\newcommand{\ese}{\end{subequations}}
\newcommand{\ba}{\begin{eqnarray}}
\newcommand{\ea}{\end{eqnarray}}
\newcommand{\bea}{\begin{eqnarray}}
\newcommand{\eea}{\end{eqnarray}}

\begin{document}

\title{Junctions, strings, clocks and gravitational memory in three dimensional dS space}

\author[1]{Avik Chakraborty,}
\emailAdd{avik.phys88@gmail.com}
\affiliation[1]{Departamento de Ciencias F\'isicas, Facultad de Ciencias Exactas, Universidad Andres Bello,
Sazi\'e 2212, Piso 7, Santiago, Chile}
\author[2,3]{Jewel Kumar Ghosh,}
\emailAdd{jewel.ghosh@iub.edu.bd}
\affiliation[2]{Department of Physical Sciences, Independent University, Bangladesh (IUB), Bashundhara RA, Dhaka 1229, Bangladesh}
\affiliation[3]{Center for Computational and Data Sciences (CCDS), Independent University, Bangladesh, Dhaka 1229, Bangladesh}
\author[4,5]{Mart\'in Molina,}
\emailAdd{martinmolinaramos95@gmail.com}
\affiliation[4]{Departamento de F\'{\i}sica, Universidad T\'{e}cnica Federico Santa Mar\'{\i}a,
Casilla 110-V, Valpara\'{\i}so, Chile,}
\author[5]{Ayan Mukhopadhyay,}
\emailAdd{ayan.mukhopadhyay@pucv.cl}
\affiliation[5]{Instituto de F\'{\i}sica, Pontificia Universidad Cat\'{o}lica de Valpara\'{\i}so,
Avenida Universidad 330, Valpara\'{\i}so, Chile}

\date{\today}

\abstract{We show that non-trivial stringy excitations in Lorentzian three dimensional de Sitter spacetime can be created self-consistently from gravitational memory in the infinite past. In addition to demonstrating that the Nambu-Goto equations for the string emerge from the two-way gravitational junction conditions, we establish the existence of well-behaved solutions corresponding to transient fluctuations of a closed string about the equator which are both borne out of and dissolve to distinct gravitational memory in the infinite past and future, respectively. The memory at infinite past, which uniquely characterizes such a solution, is a single function giving the relative angular shift at the junction gluing two two-dimensional hemispheres. This reveals that a clock dynamically emerges in the presence of a gravitational junction without the need of any external observer. 

We also show that our results generalize to the $n$-way gravitational junctions with $n\geq 3$, which are captured by Nambu-Goto-Monge-Amp\`{e}re equations for coupled $n-1$ strings -- these degrees of freedom exist even in the tensionless limit. Furthermore, for $n\geq 3$, $n-1$ correlated clocks dynamically emerge without the need of external observers in the tensionless limit, revealing a novel feature of pure three-dimensional gravity.}

\maketitle

\section{Introduction} 
Understanding quantum gravity in de Sitter (dS) spacetime remains deeply challenging. One of the approaches to formulate quantum gravity in $D+1$-dimensional de Sitter spacetime (dS$_{D+1}$) is to construct initial states with Euclidean gravitational path integral over compact geometries with $S^D$ as the boundary \cite{Hartle:1983ai,Halliwell:2018ejl} (see \cite{Lehners:2023yrj} for a review), and this is amenable to holographic interpretation \cite{Strominger:2001pn,Anninos:2011jp,Maldacena:2024uhs} in terms of a conformal field theory (CFT) living in $\mathcal{I}^+$, the infinite future boundary (S$^D$)  of dS$_{D+1}$. Some concrete examples of dS/CFT correspondence are in \cite{Anninos:2011ui,Maldacena:2019cbz,Cotler:2019nbi,Hikida:2021ese,Verlinde:2024znh}. However, such a Euclidean path integral approach needs modifications due to both technical issues \cite{Lehners:2023yrj,Feldbrugge:2017kzv} and also because it predicts a very flat Universe which is inconsistent with our observations \cite{Maldacena:2024uhs}. Other proposals involve formulating the origin of dS space as quantum tunneling \cite{Vilenkin:1982de,Linde:1998gs,Vilenkin:2018dch}, or studying quantum evolution with the Lorentzian path integral \cite{Feldbrugge:2017kzv,Feldbrugge:2017mbc}. 

More recently, it has been argued that the Hilbert space of quantum gravity in closed universes like dS is one-dimensional unless external observers are introduced for setting up clocks \cite{Marolf:2020xie,McNamara:2019rup,Usatyuk:2024mzs,Balasubramanian:2023xyd,Harlow:2025pvj,Abdalla:2025gzn,Akers:2025ahe,Nomura:2025whc,Chen:2025fwp,Aguilar-Gutierrez:2025hty,Wei:2025guh,Antonini:2025ioh}. Therefore, any consistent semiclassical or relational formulation of quantum gravity in de Sitter space should address the dynamical mechanism by which quantum reference frames are established \cite{Page:1983uc,Giacomini:2017zju,Hoehn:2019fsy,delaHamette:2020dyi}. These frames provide the clock and reference variables needed for a relational description of observables, and can then be used to construct gravitationally dressed, diffeomorphism-invariant observables in semiclassical quantum gravity \cite{Giddings:2005id,Tambornino:2011vg,Hartle:2015vfa,Chataignier:2020fys,Chandrasekaran:2022cip} (see also \cite{Geng:2024dbl} for related discussions). 

Our present work is motivated by whether extended objects like strings, which are expected to be fundamental degrees of freedom in a quantum theory of gravity, can provide a natural dynamical mechanism for defining quantum reference frames in de Sitter space. The fundamental role of strings in defining quantum gravity is that de Sitter space without a finite-dimensional Hilbert space has been emphasized earlier in \cite{Witten:2001kn}. Here, we examine this question in classical dS gravity in three dimensions, using the fundamental result that the non-linear Nambu-Goto equations of the string emerge from the gravitational junction conditions in three dimensional Einstein spacetimes \cite{Banerjee:2024sqq}.\footnote{This does not generalize for 2-way junctions to higher dimensions, as shown in \cite{Banerjee:2024sqq}. However, it is possible to have non-trivial matter-like vibrations emerging from the junction conditions even in the tensionless limit for $n$-way junctions in higher dimensions with $n\geq 3$, as discussed in \cite{Chakraborty:2025jtj}.} A generalization of this result, stating that gravitational $n$-way junctions gluing $n$ three dimensional Einstein manifolds can be described by $n-1$ strings following non-linear Nambu-Goto-Monge-Amp\`{e}re equations, has also been established in \cite{Chakraborty:2025jtj}. This generalization has the remarkable feature that even in the tensionless limit, infinite degrees of freedom survive for $n\geq 3$, demonstrating that matter-like vibrations can arise even from pure gravity. Therefore, it is natural to investigate whether the multi-way gravitational junctions in dS$_3$ can provide mechanisms of setting up dynamical reference frames even in pure gravity.

In this work, we show that although not all vibrations of the closed string correspond to well-behaved (non multi-valued) solutions of the junction conditions in dS$_3$, all classical transient vibrations of the closed string about the equator that decay both in the far past and future do correspond to well-behaved solutions of the junction conditions. Employing an early time expansion, we are also able to show the generalization of this result to $n$-way gravitational junctions in dS$_3$ for $n\geq 3$ where we need to consider the transient vibrations of $n-1$ strings coincident at the equator when coupled via Nambu-Goto-Monge-Amp\`{e}re equations.

Remarkably, we find that all such well-behaved solutions of the $n$-way gravitational junction in dS$_3$ corresponding to transient vibrations of the closed string about the equator are uniquely encoded by $n-1$ functions describing gravitational memory in the infinite past, $\mathcal{I}^-$.  These functions describing gravitational memory at $\mathcal{I}^-$ are just the $n-1$ independent relative shifts of the angular coordinate at the junction between pairs of hemispheres glued at the junction, which is located at the equator at $\mathcal{I}^-$. The transient vibrations of the strings dissolve to gravitational memory in the infinite future, $\mathcal{I}^+$, which uniquely encode the stringy vibrations as well. However, the gravitational memory at $\mathcal{I}^+$ is more complex, and we postpone a dS/CFT interpretation to the future.

Our results imply the emergence of correlated $n-1$ clocks at each point on the $n$-way gravitational junction from the junction conditions, which for a particular choice of worldsheet gauge are invariant under physical (proper) spacetime diffeomorphisms that vanish at the junction. These clocks are the $n-1$ independent relative time-shifts across any pair of dS$_3$ manifolds glued at the junction. Remarkably, for $n\geq 3$, these correlated $n-1$ clocks emerge from non-trivial solutions of the gravitational junction even in the tensionless limit (when all the isometries of de Sitter are broken), establishing a novel feature of pure three-dimensional gravity.

The rest of this paper is structured as follows. In Sec.~\ref{Sec:cl_string} we describe the solutions of the Nambu-Goto equations in dS$_3$ with special emphasis on transient vibrations of the closed string. In Sec. \ref{Sec:two_jn}, we describe the set-up of the 2-way gravitational junction, and in Sec. \ref{Sec:pert} we construct generic solutions of the junction conditions perturbatively in the tension, showing that those solutions which correspond to the transient vibrations of the closed string about the equator are well-behaved. In Sec. \ref{Sec:nway}, we describe the multi-way junctions in dS$_3$ and the perturbative expansion of the junction conditions, showing the emergence of the Nambu-Goto-Monge-Amp\`{e}re equations. In Sec. \ref{Sec:EM}, we show that all well-behaved solutions of the gravitational junction conditions corresponding to the transient stringy modes are encoded in gravitational memory involving the relative shifts of the angular coordinate at the junction in the infinite past. We also demonstrate the dynamical emergence of clocks. We conclude with a discussion and outlook in Sec.~\ref{Sec:conclusions}. 

\section{Classical strings in dS$_3$ space} \label{Sec:cl_string}

The dS$_3$ spacetime is a solution of pure Einstein's gravity in three dimensions with positive cosmological constant $\Lambda = L^{-2}$. Its line element is
\begin{equation} \label{Eq:dsmetric}
ds^2=-dt^2+L^2 \cosh^2 \left(\frac{t}{L}\right) \left( d\theta^2+\sin^2\theta d\phi^2 \right)\,
\end{equation}
describing a sphere ($S^2$) contracting from infinite volume to a minimal volume of $4\pi L^2$ during time $-\infty < t \leq 0$ and then expanding again for time $0<t < \infty$. 
For convenience, we set $L=1$.

Let us consider a closed string in dS$_3$ spacetime with its worldsheet coordinates $\tau$ and $\sigma$ are fixed by the gauge choice: $\tau = t$ and $\sigma = \phi$. Its embedding in dS$_3$ is then given by the hypersurface
\begin{equation} \label{Eq:embed_NG}
   \Sigma_{\rm NG}: \,\, t=\tau\,, \ \phi=\sigma\,, \ \theta=f_{\rm NG}(\tau,\sigma)\,.
\end{equation}
The Nambu-Goto equation for extremization of the worldsheet area of $\Sigma_{\rm NG}$ takes the explicit form
\begin{align} \label{Eq:NG_ds}
& \sin (2f_{\rm NG}) - 2f_{\rm NG}'' + 3\,\dot{f}_{\rm NG}\sin^2 (f_{\rm NG}) \sinh 2\tau  - 4\, f_{\rm NG}'\, \dot{f}_{\rm NG}\, \dot{f}_{\rm NG}' \cosh^2 \tau \notag \\&-\dot{f}_{\rm NG}^2\cosh^2 \tau \Big[\sin (2f_{\rm NG}) - 2f_{\rm NG}'' + 4\,\dot{f}_{\rm NG} \sinh \tau \cosh \tau \sin^2 (f_{\rm NG}) \Big] 
\notag\\& + 2\, \ddot{f}_{\rm NG} \sin^2 (f_{\rm NG}) \cosh^2 \tau + 2\,{f^{\prime 2}_{\rm NG}}\Big[ 2\cot (f_{\rm NG})+ \cosh \tau \big(3\,\dot{f}_{\rm NG}\sinh \tau + \ddot{f}_{\rm NG}\cosh \tau \big)\Big] = 0,
\end{align}
where the dot and prime denote derivative with respect to $\tau$ and $\sigma$, respectively.

We readily note that $f_{\rm NG} = \pi/2$ corresponding to the closed string being on an equator of the sphere for all times is an exact solution of the Nambu-Goto equation. A more general class of solutions relevant for our present discussion are the perturbations of this exact solution which take the form
\begin{equation} \label{Eq:NG_pert}
    f_{\rm NG}(\tau,\sigma)=\frac{\pi}{2}+\epsilon\, f_1(\tau,\sigma) +\epsilon^3\, f_3(\tau,\sigma) +\mathcal{O}(\epsilon^5)
\end{equation}
with $\epsilon$ an infinitesimal parameter. Note that the perturbation can be assumed to be odd in $\epsilon$ as \eqref{Eq:NG_ds} has only terms which are odd in 
$f_{\rm NG}$.

At linear order in $\epsilon$,  \eqref{Eq:NG_ds} is of the form
\begin{equation} \label{Eq:NG_lin}
 f_1 + f_1'' 
- \cosh \tau \left(3\,\dot{f}_1 \sinh \tau + \ddot{f}_1 \cosh \tau \right) \, =0.
\end{equation}
This equation has the general solution of the type
\begin{equation}\label{Eq:sol-gen}
    f_1(\tau,\sigma) = \kappa_0(\tau)+ \sum_{n=1}^\infty\left[\kappa_{n}^1(\tau) \cos (n\sigma) + \kappa_{n}^2(\tau) \sin (n\sigma) \right].
\end{equation}
Each $\kappa_n(\tau)$ for $n>0$ has two types of solutions: (i) \textit{transient} modes which decay at $\vert\tau\vert \rightarrow \infty$, and (ii) \textit{persistent} modes which 
become constants at $\vert\tau\vert \rightarrow \infty$. As for example,
\begin{equation}\label{Eq:sol-pert}
    \kappa_{2}^{1,2}(\tau) = \text{sech}^3 \tau \left[ \mathcal{A}_{1,2} + \frac{\mathcal{B}_{1,2}}{12} \left(9 \sinh \tau + \sinh 3\tau \right) \right],
\end{equation}
where the $ \mathcal{A}_{1,2}$ modes are transient and $ \mathcal{B}_{1,2}$ modes are persistent. The general solution of $\kappa_0(\tau)$ is
\begin{equation}
    \kappa_0(\tau) = \mathcal{C}\tanh \tau+ \mathcal{D}\Big[{\rm sech}\, \tau + \arctan(\sinh \tau) \tanh \tau\Big].
\end{equation}
We readily note that $\kappa_0(\tau)$ cannot decay both at $\tau \rightarrow \pm\infty$ implying that there is no transient homogeneous mode. 

We also find that the higher order perturbative corrections do not change the transient nature of the solution if we restrict 
to transient modes of $\kappa_{n}^{1,2}$ only at the first order for $n>0$. This implies the existence of transient solutions of the full non-linear equation \eqref{Eq:NG_ds} in which the closed string has only a transient
vibration and settles to the equator at $\vert \tau\vert \rightarrow \infty$.

To illustrate this general feature, we present the case of $n=2$ for which the solution at the linear order has $\kappa_{2}^{1,2}(\tau)$ of the form as given in \eqref{Eq:sol-pert}. Let us retain only the transient modes ($ \mathcal{A}_{1,2}$) by setting the persistent modes $ \mathcal{B}_{1,2} =0$. In this case, the Nambu-Goto equation amounts to:
\begin{equation}\label{Eq:f3}
f_3 + f^{\prime\prime}_3 - \cosh \tau \big( 3\dot{f}_3 \sinh \tau + \ddot{f}_3 \cosh \tau \big)\, =\mathcal{S}
\end{equation}
with
\begin{align} \label{Eq:NG_source}
\mathcal{S} =& \frac{1}{24} \text{sech}^9 \tau (\mathcal{A}_1 \cos 2\sigma + \mathcal{A}_2 \sin 2\sigma) \Big[ (\mathcal{A}_1^2 + \mathcal{A}_2^2) (641 - 918 \cosh 2\tau + 81 \cosh 4\tau) +  \notag \\
&(\mathcal{A}^2_1 - \mathcal{A}^2_2) \cos 4\sigma (257 - 54 \cosh 2\tau + 81 \cosh 4\tau) + 2 \mathcal{A}_1 \mathcal{A}_2 (257 - 54 \cosh 2\tau +  \notag \\
&81 \cosh 4\tau) \sin 4\sigma \Big],
\end{align}
at $\mathcal{O}(\epsilon^3)$. We can check that transient solutions of $f_3$ exist which vanish at $\vert\tau\vert \rightarrow\infty$. This feature continues to higher orders in the perturbative expansion.

In what follows, we will show that, remarkably, the transient solutions of the string can be created self-consistently from gravitational memory in the infinite past, and these also dissolve to 
a gravitational memory in the infinite future.

\section{Gravitational $2$-way junction} 
\label{Sec:two_jn}
The solution of the junction conditions can be found by adapting the methodology of \cite{Banerjee:2024sqq} to de Sitter space. Consider two \textit{identical} copies $\mathcal{M}_{1,2}$ of a locally dS$_3$ manifold $\mathcal{M}$, each of which is divided into two halves, northern ($N$) and southern ($S$) by codimension-$1$ hypersurfaces $\Sigma_{1,2}$.  A gravitational junction $\Sigma$ is constructed by gluing one of the fragments of $\mathcal{M}_1$ to one of the $\mathcal{M}_2$, which we denote as $\mathcal{M}_{i \alpha_i}$, with $i=1,2$ and the corresponding $\alpha_i = N,S$. The full spacetime $\widetilde{\mathcal{M}}$ together with the junction $\Sigma$ is formed by identifying the points on $\Sigma_{1,2}$. Therefore, $\Sigma_{1,2}$ should be considered as the images of the junction $\Sigma$ in $\mathcal{M}_{1,2}$, respectively. This \textit{identification} of the points of $\Sigma_i$ and the \textit{embeddings} of $\Sigma_i$ in $\mathcal{M}_i$ should satisfy the gravitational junction conditions at $\Sigma$.

\begin{figure}[ht]
    \centering
    \tikzset{
    pattern size/.store in=\mcSize, 
    pattern size = 5pt,
    pattern thickness/.store in=\mcThickness, 
    pattern thickness = 0.3pt,
    pattern radius/.store in=\mcRadius, 
    pattern radius = 1pt}
    \makeatletter
    \pgfutil@ifundefined{pgf@pattern@name@_fl5un0t35}{
    \pgfdeclarepatternformonly[\mcThickness,\mcSize]{_fl5un0t35}
    {\pgfqpoint{0pt}{0pt}}
    {\pgfpoint{\mcSize+\mcThickness}{\mcSize+\mcThickness}}
    {\pgfpoint{\mcSize}{\mcSize}}
    {
    \pgfsetcolor{\tikz@pattern@color}
    \pgfsetlinewidth{\mcThickness}
    \pgfpathmoveto{\pgfqpoint{0pt}{0pt}}
    \pgfpathlineto{\pgfpoint{\mcSize+\mcThickness}{\mcSize+\mcThickness}}
    \pgfusepath{stroke}
    }}
    \makeatother

    \tikzset{
    pattern size/.store in=\mcSize, 
    pattern size = 5pt,
    pattern thickness/.store in=\mcThickness, 
    pattern thickness = 0.3pt,
    pattern radius/.store in=\mcRadius, 
    pattern radius = 1pt}
    \makeatletter
    \pgfutil@ifundefined{pgf@pattern@name@_3pcaj0uq4}{
    \pgfdeclarepatternformonly[\mcThickness,\mcSize]{_3pcaj0uq4}
    {\pgfqpoint{0pt}{0pt}}
    {\pgfpoint{\mcSize+\mcThickness}{\mcSize+\mcThickness}}
    {\pgfpoint{\mcSize}{\mcSize}}
    {
    \pgfsetcolor{\tikz@pattern@color}
    \pgfsetlinewidth{\mcThickness}
    \pgfpathmoveto{\pgfqpoint{0pt}{0pt}}
    \pgfpathlineto{\pgfpoint{\mcSize+\mcThickness}{\mcSize+\mcThickness}}
    \pgfusepath{stroke}
    }}
    \makeatother

    \tikzset{every picture/.style={line width=0.75pt}} 

    \begin{tikzpicture}[scale=0.8, every node/.style={scale=0.6}, x=0.75pt,y=0.75pt,yscale=-1,xscale=1]

    \draw  [line width=2.25]  (81,127.5) .. controls (81,85.25) and (115.25,51) .. (157.5,51) .. controls (199.75,51) and (234,85.25) .. (234,127.5) .. controls (234,169.75) and (199.75,204) .. (157.5,204) .. controls (115.25,204) and (81,169.75) .. (81,127.5) -- cycle ;
    \draw  [line width=2.25]  (355,120.5) .. controls (355,78.25) and (389.25,44) .. (431.5,44) .. controls (473.75,44) and (508,78.25) .. (508,120.5) .. controls (508,162.75) and (473.75,197) .. (431.5,197) .. controls (389.25,197) and (355,162.75) .. (355,120.5) -- cycle ;
    \draw    (81,127.5) .. controls (121,97.5) and (197.5,99.17) .. (234,127.5) ;
    \draw  [pattern=_fl5un0t35,pattern size=6pt,pattern thickness=0.75pt,pattern radius=0pt, pattern color={rgb, 255:red, 0; green, 0; blue, 0}] (81,127.5) .. controls (83.17,130.17) and (97.84,137.25) .. (133.5,135.17) .. controls (169.16,133.08) and (185.74,137.06) .. (191.83,136.83) .. controls (197.92,136.6) and (233.63,120.9) .. (234,127.5) .. controls (234.37,134.1) and (224.5,172.17) .. (207.5,186.17) .. controls (190.5,200.17) and (174.82,204.69) .. (154.5,203.83) .. controls (134.18,202.98) and (117.5,191.83) .. (103.83,183.17) .. controls (90.17,174.5) and (78.83,124.83) .. (81,127.5) -- cycle ;
    \draw  [fill={rgb, 255:red, 15; green, 12; blue, 222 }  ,fill opacity=0.99 ] (187.5,136.83) .. controls (187.5,135.64) and (188.47,134.67) .. (189.67,134.67) .. controls (190.86,134.67) and (191.83,135.64) .. (191.83,136.83) .. controls (191.83,138.03) and (190.86,139) .. (189.67,139) .. controls (188.47,139) and (187.5,138.03) .. (187.5,136.83) -- cycle ;
    \draw    (355,120.5) .. controls (395,90.5) and (460,91.17) .. (508,120.5) ;
    \draw  [pattern=_3pcaj0uq4,pattern size=6pt,pattern thickness=0.75pt,pattern radius=0pt, pattern color={rgb, 255:red, 0; green, 0; blue, 0}] (392.67,54.83) .. controls (402.96,49.69) and (421.59,43.01) .. (438.67,44.83) .. controls (455.74,46.66) and (472.9,57.8) .. (476,58.83) .. controls (479.1,59.87) and (511,81.83) .. (508,120.5) .. controls (505,159.17) and (476.53,129.09) .. (455,127.17) .. controls (433.47,125.25) and (426.51,135.08) .. (405.67,131.83) .. controls (384.83,128.59) and (355.12,133.18) .. (355,120.5) .. controls (354.88,107.82) and (356.41,99.7) .. (363,86.5) .. controls (369.59,73.3) and (382.38,59.98) .. (392.67,54.83) -- cycle ;
    \draw  [fill={rgb, 255:red, 222; green, 12; blue, 41 }  ,fill opacity=0.99 ] (420.17,132.5) .. controls (420.17,131.3) and (421.14,130.33) .. (422.33,130.33) .. controls (423.53,130.33) and (424.5,131.3) .. (424.5,132.5) .. controls (424.5,133.7) and (423.53,134.67) .. (422.33,134.67) .. controls (421.14,134.67) and (420.17,133.7) .. (420.17,132.5) -- cycle ;
    \draw    (249.2,127.78) -- (342.4,127.02) ;
    \draw [shift={(345.4,127)}, rotate = 179.54] [fill={rgb, 255:red, 0; green, 0; blue, 0 }  ][line width=0.08]  [draw opacity=0] (8.93,-4.29) -- (0,0) -- (8.93,4.29) -- cycle    ;
    \draw [shift={(246.2,127.8)}, rotate = 359.54] [fill={rgb, 255:red, 0; green, 0; blue, 0 }  ][line width=0.08]  [draw opacity=0] (8.93,-4.29) -- (0,0) -- (8.93,4.29) -- cycle    ;

    \draw (147.83,60.57) node [anchor=north west][inner sep=0.75pt]    {$N$};
    \draw (153.17,168.23) node [anchor=north west][inner sep=0.75pt]    {$S$};
    \draw (57.17,117.9) node [anchor=north west][inner sep=0.75pt]    {$\Sigma _{1}$};
    \draw (180.5,113.23) node [anchor=north west][inner sep=0.75pt]    {$P_{1}$};
    \draw (422.33,65.23) node [anchor=north west][inner sep=0.75pt]    {$N$};
    \draw (427.67,160.9) node [anchor=north west][inner sep=0.75pt]    {$S$};
    \draw (516.33,111.9) node [anchor=north west][inner sep=0.75pt]    {$\Sigma _{2}$};
    \draw (400.67,135.23) node [anchor=north west][inner sep=0.75pt]    {$P_{2}$};
    \draw (278,97.2) node [anchor=north west][inner sep=0.75pt]    {glue};

    \end{tikzpicture}

    \caption{A two-way junction formed by gluing dS$_3$ manifolds $\mathcal{M}_1$ and $\mathcal{M}_2$. The points $P_1$ and $P_2$ on $\Sigma_1$ on $\Sigma_2$, respectively are  identified. The shaded portions of $\mathcal{M}_1$ and $\mathcal{M}_2$ are discarded  before gluing.}
    \label{fig:setup}
\end{figure}
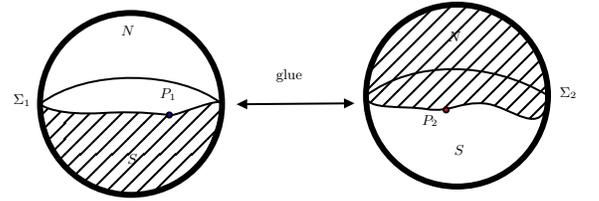

Since both copies inherit the coordinate charts of $\mathcal{M}$ with line element \eqref{Eq:dsmetric} (and $L$ set to 1), the fragments $\mathcal{M}_{i \alpha_i}$ inherit coordinates $(t_i$, $\theta_i$, $\phi_i) $.  As shown in Fig. \ref{fig:setup}, we identify unique points $P_i$ in $\Sigma_i$  to each point $P$ in $\Sigma$ with coordinates $(\tau,\sigma)$. We fix the freedom of choosing these worldsheet coordinates $(\tau,\sigma)$ by imposing the gauge fixing condition
\begin{equation} \label{Eq:WSgauge2-way}
     \tau(P)=\frac{t_1(P_1)+t_2(P_2) }{2}\,, \ \sigma(P)=\frac{\phi_1(P_1)+\phi_2 (P_2)}{2}\,.  
\end{equation}
Therefore, the embeddings of $\Sigma_i$ in $\mathcal{M}_i$ are of the form
\begin{align} \label{Eq:embed}
    &\Sigma_{1}: t_1 = \tau + \tau_d(\tau,\sigma), \,\, \phi_1 = \sigma + \sigma_d(\tau,\sigma),\,\, \theta_{1} = \mathfrak{f}_{1}(\tau,\sigma),\nonumber\\
    &\Sigma_{2}: t_2 = \tau - \tau_d(\tau,\sigma), \,\, \phi_2 = \sigma - \sigma_d(\tau,\sigma),\,\, \theta_{2} = \mathfrak{f}_{2}(\tau,\sigma).
\end{align}
Thus, in total, we have the following four functions of the worldsheet coordinates, namely
\begin{equation} \label{Eq:varbs}
    \tau_d = \frac{t_1 - t_2}{2}, \,\ \sigma_d = \frac{\phi_1 - \phi_2}{2},\, \ \theta_{s} = \frac{\mathfrak{f}_1 + \mathfrak{f}_2}{2},\,\,\theta_{d} = \frac{\mathfrak{f}_1 - \mathfrak{f}_2}{2},
\end{equation}
which should be obtained by solving the gravitational junction conditions.

The full gravitational action is
\begin{align}\label{Eq:bulk-action}
   & S=\frac{1}{16\pi G_N}\int_{\mathcal{\widetilde{\mathcal{M}}}}d^3x \sqrt{-g}(R - 2\Lambda)+T_0\int_{\Sigma}{\rm d}\tau{\rm d}\sigma\,\sqrt{-\gamma}\nonumber\\&\quad+{\rm GHY \ terms}\,,
\end{align}
where the bulk metric $g$ is the \textit{only} degree of freedom, GHY is the Gibbons-Hawking-York boundary terms, and $T_0$ is the tension of the string that constitutes the junction. The variation of the action away from the junction $\Sigma$ implies that $\mathcal{M}_{i \alpha_i}$ are Einstein manifolds. The action \eqref{Eq:bulk-action} assumes that first junction condition that states that the induced metric is continuous at the junction, i.e.
\begin{equation}\label{Eq:hcont2way}
    \gamma_{1,\mu\nu}(\tau,\sigma)= \gamma_{2,\mu\nu}(\tau,\sigma) = \gamma_{\mu\nu}(\tau,\sigma),
\end{equation}
which defines the worldsheet metric $\gamma$. Varying the action \eqref{Eq:bulk-action} with respect to $g$ at the junction $\Sigma$ yields
\begin{equation}\label{Eq:Kdisc2way}
    \sum_{i=1}^2 (-1)^{s(\alpha_i)}\left(K_{i,\mu\nu} - K_i\,\gamma_{i,\mu\nu}\right) = 8\pi G_N T_0 \gamma_{\mu\nu},
\end{equation}
with $s(\alpha_i) = 0,1$ if $\alpha_i = N,S$ respectively. $K_{i,\mu\nu}$ are the extrinsic curvatures of $\Sigma_i$ in $\mathcal{M}_{i\alpha_i}$ and $K_i = \gamma^{\mu\nu}K_{i,\mu\nu}$, respectively. The bulk diffeomorphism symmetry implies that the divergence of \eqref{Eq:Kdisc2way} vanishes. Therefore, we obtain only one independent equation from \eqref{Eq:Kdisc2way} that together with \eqref{Eq:hcont2way} give four equations which determine the four variables \eqref{Eq:varbs}. In what follows, we will glue $\mathcal{M}_{1,N}$ and $\mathcal{M}_{2,S}$ unless mentioned explicitly.

It has been shown in \cite{Banerjee:2024sqq} that the solutions of a gravitational junction gluing two identical locally flat or locally AdS three dimensional spacetimes are in one-to-one correspondence with the solutions of the non-linear Nambu-Goto equation in that spacetime. In the following section, we will show that this result extends to a generic solution of the junction conditions in dS$_3$ spacetime. Specifically,  
\begin{enumerate}
    \item the hypersurface $$\Sigma_{NG}: t=\tau, \,\, \phi=\sigma,\,\, \theta = \theta_s(\tau,\sigma)$$in $\mathcal{M}$, whose embedding is the average of $\Sigma_1$ and $\Sigma_2$, corresponds to a solution of the non-linear Nambu-Goto equations for a worldsheet in the background metric \eqref{Eq:dsmetric} when the tension $T_0$ vanishes, and
    \item $\theta_s$ is the \textit{only} degree of freedom implying that $\tau_d$, $\sigma_d$ and $\theta_d$ are completely determined as functions of $\tau$ and $\sigma$ for any given choice of the solution of the Nambu-Goto equation corresponding to $\theta_s$.
\end{enumerate}
For a junction formed by gluing $\mathcal{M}_{1,N}$ and $\mathcal{M}_{2,N}$, $\theta_s$ is exchanged with $-\theta_d$. In this case, $$\Sigma_{NG}: t=\tau, \,\, \phi=\sigma,\,\, \theta = \theta_d(\tau,\sigma)$$ corresponds to the hypersurface which follows the non-linear Nambu-Goto equation in $\mathcal{M}$.

\section{Perturbative Analysis}
\label{Sec:pert}
The general solutions of the junction conditions can be constructed perturbatively. Assuming that the dimensionless tension $\lambda=8\pi G_N T_0 L$ is $\mathcal{O}(\epsilon)$, we can solve the junction conditions perturbatively in $\epsilon$. (Recall that we also set $L=1$.) The systematic expansions of the four variables are
\begin{align} \label{Eq:pet-exp}
    &\tau_{d}=\sum_{k=1}^\infty \epsilon^k \tau_{d,k},\,\, \sigma_{d}=\sum_{k=1}^\infty \epsilon^k \sigma_{d,k},\,\,
    \theta_{d}=\sum_{k=1}^\infty \epsilon^k \theta_{d,k},\nonumber\\ &\theta_{s}=\frac{\pi}{2} +\sum_{k=1}^\infty \epsilon^k \theta_{s,k}.
\end{align}
At the zeroth order, the junction conditions, are of course, trivially satisfied.

We solve the junction conditions order by order in $\epsilon$.

\paragraph{First order:} At the first order in $\epsilon$, the conditions \eqref{Eq:hcont2way} for the continuity of the induced metric determine $\tau_{d,1}$ and $\sigma_{d,1}$. Explicitly, \eqref{Eq:hcont2way} reduce to 
\begin{align}
    & \dot{\tau}_{d,1}=0,\nonumber \\
    & \tau_{d,1}'-{\dot{\sigma}_{d,1}\cosh^2\tau }=0,\nonumber \\
    & \sigma_{d,1}'+ \tau_{d,1}\tanh\tau=0. 
\end{align}
The solutions of $\tau_{d,1}$ and $\sigma_{d,1}$ are
\begin{align}
    & \tau_{d,1}(\tau,\sigma)= \alpha_1 \sin\sigma+\alpha_2 \cos\sigma,\nonumber \\
    & \sigma_{d,1}(\tau,\sigma)=\alpha_3+ \tanh\tau \left( \alpha_1\cos\sigma-\alpha_2\sin\sigma \right). 
\end{align}
These solutions with three constant \textit{rigid parameters} $\alpha_i$, $i =1,2,3$, represent worldsheet isometries. At the zeroth order, the induced metric at the junction is dS$_2$.The Killing vectors of dS$_2$ are
\begin{align}\label{Eq:hcont2way1}
& l_1=\sin\sigma \,\partial_\tau+ \tanh \tau\cos\sigma\, \partial_\sigma, \nonumber\\
& l_2=\cos\sigma\,\partial_\tau- \tanh\tau\sin\sigma\, \partial_\sigma,\nonumber \\
& l_3=\partial_\sigma.
\end{align}
We readily note that $\alpha_1$, $\alpha_2$ and $\alpha_3$ represent the shifts in $\tau$ and $\sigma$ under the infinitesimal worldsheet diffeomorphisms generated by the Killing vectors $l_1$, $l_2$ and $l_3$, respectively. 

At the first order, we get the following three equations from the extrinsic curvature discontinuity conditions \eqref{Eq:Kdisc2way}:
\begin{align}\label{Eq:Kdisc2way1}
     \lambda + 4\, \dot{\theta}_{d,1} \sinh \tau + 2 \, \ddot{\theta}_{d,1} \cosh \tau = 0 \,, \nonumber\\
     \dot{\theta}^\prime_{d,1} = 0 \,,\nonumber \\
 \lambda \cosh \tau -2 \left(\theta_{d,1}
+ \theta_{d,1}^{\prime\prime} \right) + \dot{\theta}_{d,1}\sinh 2 \tau  = 0 \,,
\end{align}
whose solution  general solution is 
\begin{equation} \label{Eq:first-sol-thetad1}
    \theta_{d,1}=\gamma_1 \sin\sigma + \gamma_2 \cos\sigma + \gamma_3 \tanh\tau +\frac{\lambda}{2}\,\text{sech}\,\tau \,.
\end{equation}
Out of the six isometries of dS$_3$, there are three isometries which do \textit{not} preserve the hypersurface $\theta =\pi/2$. The corresponding Killing vectors are
\begin{align}
    &L_1 = \sin\phi \,\partial_\theta + \cot\theta\cos\phi\, \partial_\phi,\nonumber\\
    &L_2 = \cos\phi \,\partial_\theta - \cot\theta\sin\phi\, \partial_\phi,\nonumber\\
    &L_3 = \tanh t\,\sin\theta\, \partial_\theta -\cos\theta \,\partial_t.
\end{align}
With the zeroth order identifications of $\phi$ with $\sigma$, and $t$ with $\tau$ we readily see that $\gamma_1$, $\gamma_2$ and $\gamma_3$ represent shifts of the hypersurface $\theta = \pi/2$ under the infinitesimal diffeomorphisms generated by the Killing vectors $L_1$, $L_2$ and $L_3$, respectively. Therefore, the three constant \textit{rigid parameters} $\gamma_i$, $i =1,2,3$, are related to spacetime isometries which displace the zeroth order hypersurface giving the embedding of the junction in each spacetime (with a relative sign between $\mathcal{M}_1$ and $\mathcal{M}_2$). By virtue of being isometries, they do not affect the extrinsic curvature of the hypersurface. 

This feature of appearance of six rigid parameters related to worldsheet and spacetime isometries is also present for junctions in three-dimensional locally flat and AdS spacetimes \cite{Banerjee:2024sqq}. Here, we will set all six of these parameters $\alpha_i =\gamma_i =0$ for simplicity. In Sec. \ref{Sec:EM}, we will discuss a justification for this choice.

Note that we do not get any equation for determining $\theta_s$ from the junction conditions at first order. However, \textit{the solutions for $\tau_{d,1}$, $\sigma_{d,1}$ and $\theta_{d,1}$ exist at first order only if $\theta_{s}$ coincides with an exact solution of the Nambu-Goto equation at zeroth order.} For instance, if $\theta_s = \theta_0+\mathcal{O}(\epsilon)$ with $\theta_0\neq \pi/2$, then solutions of $\tau_{d,1}$, $\sigma_{d,1}$ and $\theta_{d,1}$ do not exist.

\paragraph{Second order:} At $\mathcal{O}(\epsilon^2)$, the metric continuity equations \eqref{Eq:hcont2way} read as follows:
\begin{align}\label{Eq:hcont-sec-eq}
\lambda \dot{\theta}_{s,1}\sinh\tau + 2 \dot{\tau}_{d,2}=0\,,\nonumber \\
\lambda \theta_{s,1}'\sinh\tau + 2 \tau_{d,2}' - 2\dot{\sigma}_{d,2} \cosh^2\tau=0\,, \nonumber\\
\lambda \theta_{s,1}
- 2 \tau_{d,2}\sinh\tau
- 2 \sigma_{d,2}'\cosh\tau=0\,.
\end{align}
From the first and third equations, which are the diagonal components of \eqref{Eq:hcont2way}, we can obtain $\tau_{d,2}$ and $\sigma_{d,2}$ explicitly in terms of $\theta_{s,1}$ up to terms that depend only on $\sigma$ or $\tau$. Substituting these forms of $\tau_{d,2}$ and $\sigma_{d,2}$ into the second equation, which is the off-diagonal component of \eqref{Eq:hcont2way}, produces a form that we denote as $\mathcal{E}$. Remarkably, 
\[
\partial_\tau\partial_\sigma(\mathcal{E}\,\text{sech}\,\tau)=0
\]
gives an equation of motion only for $\theta_{s,1}$ that is exactly the linearized Nambu-Goto equation \eqref{Eq:NG_lin} which describes fluctuations of the string about the equator with $f_1$ replaced by $\theta_{s,1}$. The general solutions of $\theta_{s,1}$ are thus of the form \eqref{Eq:sol-gen}. Furthermore, the undetermined terms in $\tau_{d,2}$ and $\sigma_{d,2}$, which are functions of  $\sigma$ or $\tau$ only, can be completely determined using $\partial_\tau\mathcal{E} =0$ and $\partial_\sigma(\mathcal{E}\,\text{sech}\,\tau)=0$ if the absence of rigid parameters, which implies the absence of the constant mode and terms proportional to $\sin \sigma$ and $\cos\sigma$ (recall the discussion above about the relationship of these terms with worldsheet isometries and see Sec. \ref{Sec:EM} for a precise discussion), is imposed. Thus, we can completely determine the solutions of $\theta_{s,1}$, $\tau_{d,2}$ and $\sigma_{d,2}$. This procedure works at higher orders in the perturbative expansion as well.

For instance, if we choose $$\theta_{s,1} = \kappa_2^1(\tau) \cos 2\sigma + \kappa_2^2(\tau) \sin 2\sigma$$ with $ \kappa_{2}^{1,2}(\tau)$ given by \eqref{Eq:sol-pert}, then
\begin{align} \label{Eq:one_theta_explicit}
\theta_{s,1}(\tau, \sigma) = \text{sech}^3 \tau \left[ \left( \mathcal{A}_1 \cos 2\sigma + \mathcal{A}_2 \sin 2\sigma \right)
  + \frac{\left( \mathcal{B}_1 \cos 2\sigma + \mathcal{B}_2 \sin 2\sigma \right)
  \left( 9 \sinh \tau + \sinh 3\tau \right)}{12}
\right].
\end{align}
Following the above procedure, we find that
\begin{align} \label{Eq:sec-vars}
\tau_{d,2}(\tau, \sigma) =& \frac{\lambda}{6} \Big[ 3\tanh^3 \tau \left(\mathcal{A}_1 \cos 2\sigma + \mathcal{A}_2 \sin 2\sigma \right) - \text{sech}\,\tau \left(3 - 2\, \text{sech}^2 \tau \right) \times \notag \\
    &\left(\mathcal{B}_1 \cos 2\sigma + \mathcal{B}_2 \sin 2\sigma \right) 
\Big], \nonumber\\
\sigma_{d,2}(\tau, \sigma) =& \frac{\lambda}{24} \Big[ 3\left(\mathcal{A}_2 \cos 2\sigma - \mathcal{A}_1 \sin 2\sigma \right) \left(1 + \tanh^2 \tau \right) 
    - 8\, \text{sech}\, \tau \tanh \tau \, \times  \notag \\ 
    &\left(\mathcal{B}_2 \cos 2\sigma - \mathcal{B}_1 \sin 2\sigma \right) - 
    9\, \text{sech}^2 \tau \left(\mathcal{A}_2 \cos 2\sigma - \mathcal{A}_1 \sin 2\sigma \right)
\Big].
\end{align}

Note that the transient modes $\mathcal{A}_i$ and the persistent modes $\mathcal{B}_i$ are $\mathcal{O}(\epsilon)$ in units where $L=1$. We readily observe that
\begin{align}
    &\lim_{\tau\rightarrow\pm\infty}\tau_{d,2} = \pm\frac{\lambda}{2}\left(\mathcal{A}_1 \cos 2\sigma + \mathcal{A}_2 \sin 2\sigma\right) +\mathcal{O}(\epsilon^2),\nonumber\\
    &\lim_{\tau\rightarrow\pm\infty}\sigma_{d,2}= \frac{\lambda}{4}\left(\mathcal{A}_2 \cos 2\sigma - \mathcal{A}_1 \sin 2\sigma\right) +  \mathcal{O}(\epsilon^2).
\end{align}
 As $\tau_d = (t_1 - t_2)/2$ and $\sigma_d = (\phi_1 - \phi_2)/2$ (see \eqref{Eq:embed}), the above implies that the transient modes exist self-consistently only if there is gravitational memory in the infinite past and future in the form of the time and angular reparametrizations at the equator. Furthermore, the transient modes can be decoded from the gravitational memory in the infinite past or future while the persistent modes are not associated with such memory at this order. These features are present for a \textit{generic} transient solution as will be explicitly shown in Sec. \ref{Sec:EM}.
 
 We note from \eqref{Eq:sec-vars} that retaining only the transient modes $\mathcal{A}_i$ renders $\tau_{d,2}$ monotonic in time implying that the increasing time shift between the two fragments of the sphere at the junction sets up a monotonic clock. We find that $\tau_{d,2}$ is generically monotonic in the far past and future when the transient excitation is practically non-existent. In Section \ref{Sec:EM}, we demonstrate that $\tau_{d,2}$ is a clock variable that emerges from gravitational memory in the infinite past, along with the transient modes of the string. This gravitational memory is encoded only in the angular shift $\lim_{\tau\rightarrow\pm\infty}\sigma_{d,2}$ at the junction in the infinite past. 

 At second order, the extrinsic curvature discontinuity conditions \eqref{Eq:Kdisc2way} are exactly as at first order (see Eq. \eqref{Eq:Kdisc2way1}) with $\theta_{d,1}$ replaced by $\theta_{d,2}$. This  allows us to set $\theta_{d,2} =0$ if we impose that $\theta_{d,2}$ vanishes in the infinite past. The initial condition that $\theta_d$ vanishes in the infinite past actually determines it uniquely to all orders in the perturbative expansion and fixes the rigid parameters related to spacetime isometries discussed above. We discuss a justification for this choice in Sec. \ref{Sec:EM}.
 
\paragraph{Third and higher orders:} At the $n$-th order in perturbative expansion, the conditions for the continuity of the induced metric \eqref{Eq:hcont2way} determine $\tau_{d,n}$, $\sigma_{d,n}$ and $\theta_{s,n-1}$. We can solve the three equations in \eqref{Eq:hcont2way} at each order $n\geq 3$ following the same algorithm adopted to solve those at the second order as these equations take the same form as \eqref{Eq:hcont-sec-eq} with $\tau_{d,2},\, \sigma_{d,2}$ and $\theta_{s,1}$ replaced by $\tau_{d,n},\, \sigma_{d,n}$ and $\theta_{s,n-1}$, respectively, with non-vanishing source terms on the right hand side for each even $n\geq 4$. Since source terms do not appear at odd $n$, we can set $\tau_{d,n}=\sigma_{d,n}=\theta_{s,n-1} =0$ for each odd $n\geq 3$ without loss of generality. In the absence of source terms, the solutions are of the same form as at the second order and can be absorbed into the constants appearing at the second order.

Similarly, at each order in the perturbative expansion, the equations for the discontinuity in the extrinsic curvature \eqref{Eq:Kdisc2way} determine $\theta_{d,n}$. Also the equations in \eqref{Eq:Kdisc2way} take the same form \eqref{Eq:Kdisc2way1} as at first order with $\theta_{d,1}$ replaced by $\theta_{d,n}$, and with non-vanishing source terms on the right hand side for each odd $n\geq 3$. In absence of source terms, we can set $\theta_{d,n} =0$ at each even $n\geq 2$ without loss of generality.  
These features parallel the same obtained in the context of junctions in locally flat or AdS three-dimensional spacetimes \cite{Banerjee:2024sqq}.

Following this general pattern, we can set
\begin{equation} \label{Eq:third-vars}
    \tau_{d,3}(\tau,\sigma) = 0\,, \ \ \sigma_{d,3}(\tau,\sigma) = 0\,, \ \ \theta_{s,2}(\tau, \sigma) = 0\,,
\end{equation}
at the third order without loss of generality while $\theta_{d,3}$ is determined by $\theta_{s,1}$. However, the generic solution is multi-valued. For instance, if $\theta_{s,1}$ is given by \eqref{Eq:one_theta_explicit}, we find that 
\begin{equation} \label{Eq:prob}
\theta_{d,3} =\frac{\lambda}{8} \Big[4 \sigma (\mathcal{A}_2 \mathcal{B}_1 - \mathcal{A}_1 \mathcal{B}_2) + \cdots\Big]\tanh \tau + \cdots\,,
\end{equation}
where the terms not shown above are non-singular and single valued functions of $\tau$ and $\sigma$. Clearly, $\theta_{d,3}$ in \eqref{Eq:prob} is multi-valued ($\sigma$ has period $2\pi$). It turns out that \textit{we find well behaved solutions of the four variables to all orders in the perturbative expansion if we retain only the transient modes}. This is obvious in \eqref{Eq:prob} if we set $\mathcal{B}_i =0$. However, this example also suggests other possibilities to obtain well-behaved solutions. We leave the investigation of more general possibilities to a future study.

After setting the persistent modes  to zero ($\mathcal{B}_1=\mathcal{B}_2=0$) in the solution \eqref{Eq:one_theta_explicit} for $\theta_{s,1}$ and demanding that 
\[
\lim_{\tau\rightarrow-\infty}\theta_{d,3}(\tau,\sigma) =0,
\]
the solution for $\theta_{d,3}$ becomes
\begin{align}
\theta_{d,3}(\tau,\sigma) &=\frac{\lambda}{7680} \, \text{sech}^7\tau \Bigg[\left(19 - 66 \cosh 2\tau + 15 \cosh 4\tau\right)\left(48(\mathcal{A}_1^2 - \mathcal{A}_2^2)\cos4\sigma + 96 \mathcal{A}_1 \mathcal{A}_2 \sin 4\sigma\right) \notag \\
&+ 5 \Bigg(732(\mathcal{A}_1^2 + \mathcal{A}_2^2) - 18\lambda^2-\left(114(\mathcal{A}_1^2 + \mathcal{A}_2^2) + 29\lambda^2\right)\cosh 2\tau \nonumber\\& 
+ \left(132(\mathcal{A}_1^2 + \mathcal{A}_2^2)-14\lambda^2\right) \cosh 4\tau + \left(18(\mathcal{A}_1^2 + \mathcal{A}_2^2) - 3\lambda^2\right) \cosh 6\tau
\notag \\
&+ 1152(\mathcal{A}_1^2 + \mathcal{A}_2^2) \arctan\left(\tanh\frac{\tau}{2}\right)\cosh^6\tau\sinh\tau
\Bigg) \Bigg]\notag\\&
+\frac{3}{16}\pi \lambda\left(\mathcal{A}_1^2 + \mathcal{A}_2^2\right) \tanh \, \tau,
\end{align}
which is indeed a well-behaved solution. Furthermore, we note that in this case
\begin{align} \label{Eq:lt_xsxa}
    \lim_{\tau \to \infty} \theta_{d,3}(\tau, \sigma) &= \frac{3}{8}\pi \lambda\left(\mathcal{A}_1^2 + \mathcal{A}_2^2\right)\,.
\end{align}
Since $\theta_d = (\theta_1 -\theta_2)/2$, this implies that the gravitational memory is also in the form of gain/loss of spatial volume at infinite future at positive/negative tension due to the presence of the gravitational junction. 

At higher orders in $\epsilon$, we recover the non-linear corrections of the Nambu-Goto equation for $\theta_s$ when $\lambda$ is taken to zero. 

For instance, after setting the persistent  $\mathcal{B}_i$ modes to zero in the solution \eqref{Eq:one_theta_explicit} for $\theta_{s,1}$, we find by solving the junction conditions for the two way junction at $\mathcal{O}(\epsilon^4)$ that $\theta_{s,3}$ satisfies the equation
\begin{align} \label{Eq:fourth_NG_correction}
\theta_{s,3} + \theta_{s,3}'' - \cosh \tau \big( 3 \,\dot{\theta}_{s,3} \sinh \tau + \ddot{\theta}_{s,3} \cosh \tau \big) = \mathcal{S} +  \frac{3\lambda^2}{8} \Big[\text{sech}^5 \tau\,(7-5\cosh 2\tau)\times  \notag \\
\left(\mathcal{A}_1 \cos 2\sigma + \mathcal{A}_2 \sin 2\sigma \right)\Big],
\end{align}
where $\mathcal{S}$ is given by \eqref{Eq:NG_source}. Clearly, $\theta_{s,3}$ satisfies the same equation \eqref{Eq:f3} as $f_3$, the first non-linear correction to the Nambu-Goto equation, when $\lambda\rightarrow 0$. Thus, the full non-linear Nambu-Goto equation is reproduced from the junction condition order by order in the perturbative expansion. Furthermore, as precisely stated in Se. \ref{Sec:EM}, the solutions corresponding to the transient vibrations of the string are well-behaved (non multi-valued).

\section{Gravitational multiway junctions and Nambu-Goto-Monge-Amp\`{e}re equations}
\label{Sec:nway}

Following \cite{Chakraborty:2025jtj}, we can construct multi-way junctions in dS$_3$ in perturbative expansion. Consider $n\geq 3$ \textit{identical} copies $\mathcal{M}_i$ (with $i=1,\cdots,n$) of  locally dS$_3$ manifold $\mathcal{M}$ \eqref{Eq:dsmetric}, each of which is again divided into two halves, $\mathcal{M}_{iN}$ and $\mathcal{M}_{iS}$, by distinct co-dimension one hypersurfaces $\Sigma_i$. A gravitational junction $\Sigma$ is formed by gluing $n$ such fragments $\mathcal{M}_{\alpha_i}$, with $\alpha_i=N,S$, resulting in the full spacetime $\widetilde{\mathcal{M}}$ that satisfies Einstein's equations along with the required junction conditions. Each point $P$ in the junction $\Sigma$ is constructed by identifying the corresponding points $P_i$ in $\Sigma_i$. Hence, each $\Sigma_i$ should be considered as the image of $\Sigma$ in the corresponding $\mathcal{M}_i$. These \textit{identifications} of the points $P_i$ of $\Sigma_i$ and the \textit{embeddings} of $\Sigma_i$ in $\mathcal{M}_i$ should satisfy the gravitational junction conditions \cite{Israel:1966rt} at $\Sigma$.  Since all the $n$ copies $\mathcal{M}_i$ inherit the coordinate charts of $\mathcal{M}$, the fragments $\mathcal{M}_{\alpha_i}$ have coordinates $(t_i,\theta_i,\phi_i)$. The embeddings of $\Sigma_i$ in $\mathcal{M}_i$ are specified by the $n$ functions
\begin{equation}\label{Eq:embed}
  \Sigma_i:\,\,  \theta_i = f_{i}(t_i, \phi_i)\,, \ \ i = 1, 2,\cdots,n\,.
\end{equation}
We form the junction $\Sigma$ by identifying the points $P_i$, with coordinates ($\tau_i,\sigma_i$), to the point $P$ in $\Sigma$, with coordinates $(\tau,\sigma)$. We fix the worldsheet gauge by choosing the coordinates $(\tau,\sigma)$ as
\begin{align}\label{Eq:WSgauge}
    \tau(P) = \frac{1}{n}\sum_{i=1}^n t_i(P_i), \quad \sigma(P) = \frac{1}{n}\sum_{i=1}^n \phi_i(P_i)\,.
\end{align}
Then we are left with $2(n-1)$ independent variables, $t_i - t_j$ and $\phi_i - \phi_j$ ($i\neq j$), which are the relative shifts of time and angle, respectively, as we move from $\mathcal{M}_{i\alpha_i}$ to $\mathcal{M}_{j\beta_j}$ across the junction $\Sigma$. Therefore, together with the $n$ embedding functions $f_i$ of $\Sigma_i$, we have in total $3n-2$ variables that completely specify the junction. The first junction condition then becomes
\begin{align}\label{Eq:hcont}
&\gamma_{\mu\nu}(\tau, \sigma) := \gamma_{1,\mu\nu}(\tau,\sigma)=\cdots = \gamma_{n,\mu\nu}(\tau,\sigma)\,,
\end{align}
and the variation of the action \eqref{Eq:bulk-action} with respect to the bulk metric at gives
\begin{equation}\label{Eq:Kdisc}
    \sum_{i=1}^n (-1)^{s(\alpha_i)}\left(K_{i,\mu\nu} - K_i\,\gamma_{i,\mu\nu}\right) = \lambda \gamma_{\mu\nu}\,
\end{equation}
at the junction with $s(\alpha_i) = 1$, if $\alpha_i = N$ and $s(\alpha_i) = -1$, if $\alpha_i = S$. The bulk diffeomorphism symmetry implies that the \textit{total} Brown-York tensor of the junction, which is the left hand side of \eqref{Eq:Kdisc}, is conserved. Therefore, we obtain only one independent equation from \eqref{Eq:Kdisc} as before, which together with $3(n-1)$ equations from \eqref{Eq:hcont} give $3n-2$ independent equations, exactly matching the number of unknown variables. 

For simplicity, we work with the \textit{folded picture} in which $\alpha_i =N$ for all $i$. This is different from the case of the two-way junction in the previous section where we chose $\alpha_1 = N$ and $\alpha_2 = S$. As mentioned before, if $\alpha_1 = \alpha_2 = N$ for the two-way junction, then we should exchange $\theta_s$ and $-\theta_d$, and therefore in this case, $\theta_d$ instead of $\theta_s$ satisfies the Nambu-Goto equation in dS$_3$ when the tension vanishes.

Following \cite{Chakraborty:2025jtj}, we define $3n-3$ independent relative shifts of the time ($\tau_{d_i}$), the azimuthal angle ($\sigma_{d_i}$), and the transverse polar coordinate ($\theta_{d_i}$), across the junction as
\begin{eqnarray}\label{Eq:xsxd}
\tau_{d_i} &=&\begin{cases}
        \frac{1}{n}(t_n-t_{i+1}) \,\, {\rm for}\,\, i = 1, \cdots, n-2\\
        \frac{1}{n}(t_n-t_1) \,\, {\rm for}\,\, i = n-1
    \end{cases}\,,\nonumber\\
\sigma_{d_i} &=&\begin{cases}
        \frac{1}{n}(\phi_n-\phi_{i+1}) \,\, {\rm for}\,\, i = 1, \cdots, n-2\\
        \frac{1}{n}(\phi_n-\phi_1) \,\, {\rm for}\,\, i = n-1
    \end{cases}\,,\nonumber\\
\theta_{d_i} &=&\begin{cases}
        \frac{1}{n}(\theta_n-\theta_{i+1}) \,\, {\rm for}\,\, i = 1, \cdots, n-2\\
        \frac{1}{n}(\theta_n-\theta_1) \,\, {\rm for}\,\, i = n-1
    \end{cases}\,.
\end{eqnarray}
Together with the average transverse coordinate
\begin{equation}
    \theta_s = \frac{1}{n}\sum_i \theta_{i},
\end{equation} 
we obtain the necessary $3n-2$ functions of $\tau$ and $\sigma$ that we need to determine. If a subset of the  $n$ fragments, $\mathcal{M}_{i\alpha_i}$ are $\mathcal{M}_{iS}$ instead of $\mathcal{M}_{iN}$, we simply reverse the sign of the transverse coordinate $\theta_i$ in the parameterization \eqref{Eq:xsxd} for the values of $i$ in this subset. 

We solve the general junction conditions perturbatively in $\lambda$ for the gluing of the $n$ fragments $\mathcal{M}_{iN}$. With $\lambda = \mathcal{O}(\epsilon)$, the systematic expansions of different variables are
\begin{align} \label{Eq:pet-exp}
    &\tau_{d_i}=\sum_{k=0}^\infty \epsilon^k \tau_{d_i,k},\,\, \sigma_{d_i}=\sum_{k=0}^\infty \epsilon^k \sigma_{d_i,k},\,\,
    \theta_{d_i}=\sum_{k=0}^\infty \epsilon^k \theta_{d_i,k},\nonumber\\ &\qquad\qquad\qquad \theta_{s}=\frac{\pi}{2} +\sum_{k=1}^\infty \epsilon^k \theta_{s,k}.
\end{align}
At the zeroth order, of course, the junction conditions are trivially satisfied.

At the first order in $\epsilon$, the solutions for $\tau_{d_i}$ and $\sigma_{d_i}$ correspond to the three worldsheet Killing vectors whereas the solution for $\theta_{s,1}$ feature the three spacetime isometries of $\mathcal{M}$ which do not preserve the zeroth order hypersurface $\theta_s = \frac{\pi}{2}$. In what follows, we set these $3n$ rigid parameters to zero, so that we choose the permutation symmetric solution at this order. Then from the extrinsic curvature discontinuity condition we find that 
\begin{equation}\label{Eq:thetassol}
    \theta_{s,1} = -\frac{\lambda }{n}\,\text{sech}\,\tau.
\end{equation}
Note that we do not get an equation for $\theta_{d_i,1}$ from the junction conditions at the first order (as in the case of the two-way junction).

At $\mathcal{O}(\epsilon^2)$, from the diagonal $\tau\tau$ and $\sigma\sigma$ components of the metric continuity conditions we can obtain $\tau_{d_i,2}$ and $\sigma_{d_i,2}$ explicitly in terms of $\theta_{d_i,2}$ up to terms which depend only on $\sigma$ or $\tau$. Substituting these forms of $\tau_{d_i,2}$ and $\sigma_{d_i,2}$ into the off-diagonal components of the metric continuity conditions yield $n-1$ equations that are schematically $\mathcal{E}_i =0$. Then $\partial_\tau\partial_\sigma(\mathcal{E}_i\,\text{sech}\,\tau)=0$ give the coupled Nambu-Goto-Monge-Amp\`{e}re equations involving \textit{only} the $n-1$ variables $\theta_{d_i,1}$. Explicitly, these coupled Nambu-Goto-Monge-Amp\`{e}re equations are
\begin{align}\label{Eq:CoupledNGEOMs}
    2\,\lambda\,\mathcal{N}_i = n\left((n-2) \mathcal{X}_i + 2\sum_{j\neq i}\mathcal{Z}_{ij} \right),
\end{align}
for $i=1,\cdots ,n-1$, with
\begin{align}\label{Eq:CNG}
   \mathcal{N}_i=\theta_{d_i,1} + \theta_{d_i,1}'' - \cosh \tau \Big( 3 \,\dot{\theta}_{d_i,1} \sinh \tau + \ddot{\theta}_{d_i,1} \cosh \tau \Big)
\end{align}
corresponding to the linearized Nambu-Goto equation for the hypersurface
\begin{equation}
    \Sigma_{NG_i}: t=\tau, \,\, \phi=\sigma,\,\, \theta = \theta_{d_i,1}(\tau,\sigma)
\end{equation}
in $\mathcal{M}$, while
\begin{align}
\mathcal{X}_i&=4\,\dot{\theta}_{d_i,1}\left(\theta_{d_i,1}+\theta_{d_i,1}'' \right) \sinh \tau \notag \\
    &\qquad\qquad- 2\left(2\,\dot{\theta}_{d_i,1}^2\sinh^2 \tau + \dot{\theta}_{d_i,1}^{'2} - \ddot{\theta}_{d_i,1}\left(\theta_{d_i,1}+\theta_{d_i,1}''-\dot{\theta}_{d_i,1}\sinh\tau\cosh\tau \right) \right)\cosh \tau
\end{align}
and
\begin{align}
\mathcal{Z}_{ij}&=4\,\dot{\theta}_{d_i,1}\dot{\theta}_{d_j,1}\sinh^2\tau\cosh\tau + 2\,\dot{\theta}_{d_i,1}'\dot{\theta}_{d_j,1}'\cosh\tau + \left(\dot{\theta}_{d_i,1}\ddot{\theta}_{d_j,1}+\ddot{\theta}_{d_i,1}\dot{\theta}_{d_j,1} \right)\sinh\tau\cosh^2\tau \notag \\
&\qquad- \left(\theta_{d_i,1}+\theta_{d_i,1}'' \right)\left(2\,\dot{\theta}_{d_j,1}\sinh\tau+\ddot{\theta}_{d_j,1}\cosh\tau \right) \notag \\
&\qquad\qquad-\left(\theta_{d_j,1}+\theta_{d_j,1}'' \right)\left(2\,\dot{\theta}_{d_i,1}\sinh\tau+\ddot{\theta}_{d_i,1}\cosh\tau \right).
\end{align}
Note that the right hand side of these equations \eqref{Eq:CoupledNGEOMs} vanish for $n=2$. Given that $\theta_{d_i}$ satisfies these equations, we can determine  $\tau_{d_i}$ and $\sigma_{d_i}$ completely from $\partial_\tau\mathcal{E}_i =0$ and $\partial_\sigma(\mathcal{E}_i\,\text{sech}\,\tau)=0$. 
 This procedure works at higher orders in the perturbative expansion also, as in the case of the two-way junction.

Due to the non-linear nature of the Nambu-Goto-Monge-Amp\`{e}re equations, finding exact solutions is harder. However, we can find solutions perturbatively in the early (late) time expansion as discussed in the next section. Remarkably, non-trivial solutions exist even in the tensionless limit implying that matter-like behavior can arise from pure gravity involving junctions gluing three or more three-dimensional spacetimes. This was also the case in the AdS spacetimes \cite{Chakraborty:2025jtj}.

We note that the $n$-way junction for $n\geq 3$ can be always reduced to the $2$-way junction. This reduction corresponds to setting $\theta_k = \theta_s$ for $k\neq i$ and $k\neq j$ (freezing the relative motions of $n-2$ of the $n$ hypersurfaces glued at the junction) which impose $n-2$ linear constraints on the $n-1$ variables $\theta_{d_i}$ so that only one linear combination of the $\theta_{d_i}$ is an independent variable. This linear combination satisfies the Nambu-Goto equation. We indeed note from \eqref{Eq:CoupledNGEOMs} that the $n-2$ constraints (imposing $\theta_k = \theta_s$ for $k\neq i$ and $k\neq j$) imply that the quadratic Monge-Amp\`{e}re terms vanish identically. As for instance, setting $\theta_1 = \theta_3 = \theta_s$ in the case of the $4$-way junction impose that $\theta_{d_1,1} = 2 \theta_{d_2,1}$ and $\theta_{d_3,1} = \theta_{d_2,1}$. In this case, the  Nambu-Goto-Monge-Amp\`{e}re equations \eqref{Eq:CoupledNGEOMs} reduce simply to the linearized Nambu-Goto equation for $\theta_{d_2,1}$. 

In the following section, we demonstrate by employing the early time expansion that solutions of the $n$-way gravitational junction corresponding to the transient vibrations of $n-1$ coincident strings at the equator exist and are well-behaved also. We also identify how such non-trivial vibrations are encoded in the gravitational memory in the infinite past.

\section{Gravitational memory as initial data and emergent clocks}\label{Sec:EM}

\subsection{The two-way junction}
We have established that generic solutions of the two-way gravitational junction in dS$_3$ correspond to transient vibrations of the string in dS$_3$ that are described by the non-linear Nambu-Goto equation. However, generic solutions of the non-linear Nambu-Goto equation do not correspond to well-behaved (non-multi-valued) solutions of the gravitational junction.  We have demonstrated that particularly the solutions of the gravitational junction in dS$_3$ corresponding to \textit{transient} vibrations of the string about the equator, which decay both in the far past and the far future, are well-behaved. This demonstration has required the absence of rigid parameters that amounted to a choice of integration constants. A pertinent question is \textit{which initial data uniquely characterizes} the well-behaved solutions of the two-way gravitational junction, and what the \textit{absence of rigid parameters} precisely means.

To identify the initial data that uniquely characterize the solutions of the gravitational junction, we can perform the early-time expansion. This expansion is analogous to the Fefferman-Graham expansion (or more general radial expansion) in AdS, which demonstrates how the boundary data characterize a generic asymptotically AdS solution of classical gravity. In the context of the solutions of the gravitational junction, it is easier to achieve both the early time expansion in dS and the radial expansion in AdS perturbatively in the tension. Particularly, we need to \textit{first} expand the equations of gravitational junction perturbatively in the tension (as has been done in the previous sections) and \textit{then} perform the early-time expansion as explained below. In fact, as shown below, it is sufficient to identify the initial data only from the second order in the perturbative expansion in the tension, as the solutions at higher orders in the tension are uniquely specified by the initial data identified at the second order. 

We recall the expansion of all the four variables in the dimensionless tension $\lambda$ given by \eqref{Eq:pet-exp} with $\lambda = \mathcal{O}(\epsilon)$ and that at the first order in the perturbative expansion we set
\begin{equation}
    \tau_{d,1} = \sigma_{d,1} = 0, \quad \theta_{d,1} = \frac{\lambda}{2}{\rm sech}\, \tau.
\end{equation}
In fact, these are the unique solutions which satisfy
\begin{equation}\label{Eq:ic-first-order}
    \lim_{\tau\rightarrow-\infty}\sigma_{d,1}(\tau,\sigma) = \lim_{\tau\rightarrow-\infty}\theta_{d,1}(\tau,\sigma) =0.
\end{equation}

At the second order in the perturbative expansion, we can solve the equations for the continuity of the induced metric \eqref{Eq:hcont-sec-eq} using the early-time expansion. We assume that at $\tau\sim -\infty$, the three variables $\tau_{d,2}$, $\sigma_{d,2}$ and $\theta_{s,1}$ can be expanded as follows:
\begin{equation}
    \tau_{d,2}(\tau,\sigma)=\sum_{k=0}^{\infty}\tau_{d,2}^{(k)}(\sigma)e^{k\tau}\,, \ \sigma_{d,2}(\tau,\sigma)=\sum_{k=0}^{\infty}\sigma_{d,2}^{(k)}(\sigma)e^{k\tau}\,, \  \theta_{s,1}(\tau,\sigma)=\sum_{k=0}^{\infty}\theta_{s,1}^{(k)} (\sigma)e^{k\tau}\,.
\end{equation}
We can readily derive from the equations for the continuity of the induced metric \eqref{Eq:hcont-sec-eq} that all the coefficients of the above expansion in terms of only
\[
\sigma_{d,2}^{(0)} = \lim_{\tau \rightarrow -\infty} \sigma_{d,2} \quad {\rm and} \quad \theta_{s,2}^{(0)} = \lim_{\tau \rightarrow -\infty} \theta_{s,2}.
\]
Explicitly,
\begin{align} \label{Eq:coeff_longtime}
    &\tau_{d,2}^{(0)} =  \sigma_{d,2}^{(0)\prime}\,, \notag \\
    &\sigma_{d,2}^{(1)} = - \lambda\theta_{s,1}^{(0)\prime}\,, \  \tau_{d,2}^{(1)} = -\lambda\left(\theta_{s,1}^{(0)}+\theta_{s,1}^{(0)\prime\prime}\right)\,, \ \theta_{s,1}^{(1)} = 0\,, \notag \\
    &\sigma_{d,2}^{(2)} = 2\sigma_{d,2}^{(0)\prime\prime}\,, \  \tau_{d,2}^{(2)} = 2\left(\sigma_{d,2}^{(0)\prime}+\sigma_{d,2}^{(0)\prime\prime\prime}\right)\,, \ \theta_{s,1}^{(2)} = -2\left(\theta_{s,1}^{(0)}+\theta_{s,1}^{(0)\prime\prime}\right)\,, \notag \\
    &\sigma_{d,2}^{(3)} = \frac{\lambda}{3}\left(\theta_{s,1}^{(0)\prime}-2\theta_{s,1}^{(0)\prime\prime\prime}\right)\,, \  \tau_{d,2}^{(3)} = \frac{\lambda}{3}\left(3\theta_{s,1}^{(0)}+\theta_{s,1}^{(0)\prime\prime}-2\theta_{s,1}^{(0)\prime\prime\prime\prime}\right)\,, \ \theta_{s,1}^{(3)} = \frac{16}{3\lambda}\left(\sigma_{d,2}^{(0)\prime}+\sigma_{d,2}^{(0)\prime\prime\prime} \right)\,, \notag \\
    &\sigma_{d,2}^{(4)} = -\frac{2}{3}\left(2\sigma_{d,2}^{(0)\prime\prime}-\sigma_{d,2}^{(0)\prime\prime\prime\prime}\right)\,, \  \tau_{d,2}^{(4)} = -\frac{2}{3}\left(5\sigma_{d,2}^{(0)\prime}+4\sigma_{d,2}^{(0)\prime\prime\prime}-\sigma_{d,2}^{(0)\prime\prime\prime\prime\prime}\right)\,, \nonumber\\ &\theta_{s,1}^{(4)} = 2\left(\theta_{s,1}^{(0)}-\theta_{s,1}^{(0)\prime\prime\prime\prime}\right), \cdots.
\end{align}
The reader can readily verify that the above results are valid for the exact solutions of $\theta_{s,1}$, $\sigma_{d,2}$ and $\tau_{d,2}$ discussed earlier, for instance those given in \eqref{Eq:one_theta_explicit} and \eqref{Eq:sec-vars}. 

We particularly note from \eqref{Eq:coeff_longtime} that
\begin{align}
      &\theta_{s,1}^{(3)} = \frac{16}{3\lambda}\left(\sigma_{d,2}^{(0)\prime}+\sigma_{d,2}^{(0)\prime\prime\prime} \right),\label{Eq:key-relation} \\ &\sigma_{d,2}^{(3)} = \frac{\lambda}{3}\left(\theta_{s,1}^{(0)\prime}-2\theta_{s,1}^{(0)\prime\prime\prime}\right)\,,
\end{align}
implies that $\sigma_{d,2}^{(0)}$ sources $\theta_{s,1}^{(3)}$, and $\theta_{s,1}^{(0)}$ sources $\sigma_{d,2}^{(3)}$. Such relationships are far more complicated and difficult to disentangle without employing the perturbative expansion in the tension prior to the early time expansion. In fact, \eqref{Eq:key-relation} explicitly shows that the $\lambda\rightarrow 0$ limit is singular, and so it is better to understand the early time behavior of the solution employing the perturbative expansion in the tension first.

In the context of the gravitational junction in AdS/CFT, it was essential to set the Dirichlet boundary condition to have an interpretation of the junction in terms of an interface in the dual CFT, so that we had only $\tau_{d,2}^{(0)}$, the time-jump across the interface. Also $\tau_{d,2}^{(0)}$ sourced the normalizable Nambu-Goto mode $x_{s,1}^{(3)}$ (with $x$, the transverse coordinate analogous to $\theta$), which has the interpretation of the \textit{displacement operator} of the interface via the interface Ward identity after the time jump was undone by a half-sided conformal transformation to establish physical coordinates that were continuous through the interface \cite{Chakraborty:2025dmc,Banerjee:2025zuw}. In the present context of the gravitational junction in dS$_3$, we recall from Sec. \ref{Sec:pert} that higher order expansions in $\epsilon$ show that we can avoid multi-valued solutions if we set
\[
\theta_{s,1}^{(0)} =0.
\]
Unlike the case of AdS, we set $\theta_{s,1}^{(0)} =0$ to obtain well-behaved solutions rather than being motivated by dS/CFT interpretations. Setting $\theta_{s,1}^{(0)} =0$, simplifies the early time expansion, giving
\begin{align} \label{Eq:coeff_longtime_2}
    &\tau_{d,2}^{(0)} =  \sigma_{d,2}^{(0)\prime}\,, \notag \\
    &\sigma_{d,2}^{(1)} = 0,, \  \tau_{d,2}^{(1)} = 0\,, \ \theta_{s,1}^{(1)} = 0\,, \notag \\
    &\sigma_{d,2}^{(2)} = 2\sigma_{d,2}^{(0)\prime\prime}\,, \  \tau_{d,2}^{(2)} = 2\left(\sigma_{d,2}^{(0)\prime}+\sigma_{d,2}^{(0)\prime\prime\prime}\right)\,, \ \theta_{s,1}^{(2)} = 0\,, \notag \\
    &\sigma_{d,2}^{(3)} =0\,, \  \tau_{d,2}^{(3)} = 0, \ \theta_{s,1}^{(3)} = \frac{16}{3\lambda}\left(\sigma_{d,2}^{(0)\prime}+\sigma_{d,2}^{(0)\prime\prime\prime} \right)\,, \notag \\
    &\sigma_{d,2}^{(4)} = -\frac{2}{3}\left(2\sigma_{d,2}^{(0)\prime\prime}-\sigma_{d,2}^{(0)\prime\prime\prime\prime}\right)\,, \  \tau_{d,2}^{(4)} = -\frac{2}{3}\left(5\sigma_{d,2}^{(0)\prime}+4\sigma_{d,2}^{(0)\prime\prime\prime}-\sigma_{d,2}^{(0)\prime\prime\prime\prime\prime}\right)\,, \ \theta_{s,1}^{(4)} = 0, \cdots
\end{align}
so that the full solution of the three variables is determined only by $\sigma_{d,2}^{(0)}$. 

We recall from our previous discussion in Sec \ref{Sec:pert} the general feature that we can set $\tau_{d,n}=\sigma_{d,n}=\theta_{s,n-1} =0$ for each odd $n\geq 3$ and $\theta_{d,n} =0$ at each even $n\geq 2$ without loss of generality. Furthermore, we obtain unique well-behaved solutions to all orders by demanding that\footnote{One can actually more generally specify that $\lim_{\tau\rightarrow -\infty} \sigma_{d,n}(\tau,\sigma) = g_{n}(\sigma)$ and $\lim_{\tau\rightarrow -\infty} \theta_{d,n}(\tau,\sigma) = \lim_{\tau\rightarrow -\infty} \theta_{s,n}(\tau,\sigma) =0$ at all orders. Our arguments do not change with this modification. We have chosen the simplest illustration of the principle that the infinite past value of $\sigma_{d,n}$ fully specifies the solutions corresponding to transient vibrations of the string about the equator.}
\begin{align}\label{Eq:ic-higher-order}
    &\lim_{\tau\rightarrow -\infty} \theta_{d,n}(\tau,\sigma) = 0 \,\, \forall \,\, {\rm odd} \, n\geq 3,\notag\\
    &\lim_{\tau\rightarrow -\infty} \sigma_{d,n}(\tau,\sigma) =\lim_{\tau\rightarrow -\infty} \theta_{s,n-1}(\tau,\sigma) =0 \,\,\forall \,\, {\rm even} \,\, n\geq2.
\end{align}

Therefore, together with the results of the first and second orders in the perturbative $\epsilon$ expansion discussed above, we can conclude that the full solution of the gravitational junction is unique and well-behaved when we set initial conditions in the far past in the form
\begin{align}\label{Eq:ic-full}
   &\lim_{\tau\rightarrow -\infty} \sigma_d(\tau,\sigma) = f(\sigma),\notag\\
   &\lim_{\tau\rightarrow -\infty} \theta_d(\tau,\sigma) = \lim_{\tau\rightarrow -\infty} \theta_s(\tau,\sigma) =0,
\end{align}
where $f(\sigma)$ is a dimensionless function (in units where the Hubble scale is unity). To be consistent with the perturbative expansion, we require that $\lambda = \mathcal{O}(\epsilon)$ and $f(\sigma)= \mathcal{O}(\epsilon^2)$, so that $\sigma_{d,2}^{(0)}$ can be identified with $f(\sigma)$.\footnote{We can allow higher $\epsilon$-order terms in $f(\sigma)$ as noted in the previous footnote.} Finally, the absence of the three rigid parameters corresponding to the worldsheet isometries implies that additionally,
\begin{equation}\label{Eq:No-rigid}
    \int _0^{2\pi}{\rm d}\sigma\ f(\sigma) = \int _0^{2\pi}{\rm d}\sigma\ f(\sigma) \cos\,\sigma =\int _0^{2\pi}{\rm d}\sigma\ f(\sigma) \sin\,\sigma = 0.
\end{equation}
These imply that the Fourier expansion $f(\sigma)$ does not have the constant mode and the harmonics proportional to $\sin\,\sigma$ and $\cos \, \sigma$. Note that the initial condition on $\theta_{d}$ itself fixes the three rigid parameters corresponding to spacetime isometries. We say that the solutions with initial conditions given in \eqref{Eq:ic-full} where $f(\sigma)$ satisfy \eqref{Eq:No-rigid} are bereft of all rigid parameters. Such solutions reduce to the full smooth dS$_3$ manifold in the tensionless limit justifying the imposition of \eqref{Eq:No-rigid}.

We have thus demonstrated that \textit{solutions of the gravitational junction in dS$_3$ which correspond to the transient vibrations of the string about the equator and reduce to the smooth dS$_3$ manifold in the tensionless limit are uniquely determined by the gravitational memory in the far past, which is encoded in a \textit{single} variable, namely $f(\sigma)$, the shift in the angular coordinate of the two hemispheres across the junction in the far past.} Such solutions are well-behaved, with $f(\sigma)$ encoding the entire gravitational solution, including all transient excitations of the stringy degree of freedom of the junction that are significant only after a long time. It is interesting to contrast this with inflation, which obliterates memory of the initial conditions.

We also note that, although $\theta_{d}$ vanishes in the far past as a result of the initial conditions \eqref{Eq:ic-full}, it does not vanish in the far future and is determined by $f(\sigma)$ as evident from \eqref{Eq:lt_xsxa}. This implies that the gravitational memory to which the transient vibration of the string dissolves to in the far future ($\mathcal{I}^+$) is more complex than the initial gravitational memory in the far past ($\mathcal{I}^-$). Given that $\theta_{d}$ does not vanish in the far future, it is not easy to interpret our solutions in terms of interfaces of a dual CFT in the dS/CFT correspondence. We postpone such an analysis to the future.

Crucially, we note that as the initial conditions are fully set by $\sigma_{d}$, $\theta_d$ and $\theta_s$ only, we can think of the clock variable $\tau_d$, the relative time between the two hemispheres, as \textit{emergent}. This implies that the \textit{gravitational junction provides a natural clock that emerges from the dynamics in the absence of any external observer.} The worldsheet gauge choice \eqref{Eq:WSgauge2-way} and the clock $\tau_d(\tau,\sigma)$ at each point on the junction are invariant under physical (proper) spacetime diffeomorphisms generated by vector fields $\xi_i^\mu$ on the 2 manifolds for which $\xi_i^t$ and $\xi_i^\phi$ vanish at the location of the junction for each $i=1,2$.\footnote{We can consistently forbid the improper combined spacetime diffeomorphisms which affect the boundary set by the junction} The rigidity conditions \eqref{Eq:No-rigid} together with the initial conditions \eqref{Eq:ic-full} ensure that these clocks cannot be undone by worldsheet and spacetime isometry transformations, which produce solutions related by the isometries of the trivial maximally symmetric solution in the tensionless limit. 

\subsection{The multi-way case}
The discussion in the previous subsection readily generalizes to the $n$-way junction with the additional feature that one obtains an infinite dimensional space of non-trivial solutions in terms of initial data even in the tensionless limit and in the absence of rigid parameters when $n\geq 3$. As discussed in Sec. \ref{Sec:nway}, such solutions are possible because the Monge-Amp\`{e}re equations that couple the strings have non-trivial solutions even in the tensionless limit, implying the emergence of matter-like degrees of freedom from pure gravity.

For the $n$-way junction, we set the initial conditions
\begin{align}\label{Eq:ic-full_n}
   &\lim_{\tau\rightarrow -\infty} \sigma_{d_i}(\tau,\sigma) := \sigma_{d_i}^{(0)} = f_i(\sigma), \ \ \forall \ \ i =1,\cdots n-1,\notag\\
   &\lim_{\tau\rightarrow -\infty} \theta_{d_i}(\tau,\sigma) = \lim_{\tau\rightarrow -\infty} \theta_s(\tau,\sigma) =0,
\end{align}
with $f_i(\sigma)$ of $\mathcal{O}(\epsilon^2)$ to be consistent with the perturbative expansion\footnote{As noted in the previous two footnotes, we can allow $f_i(\sigma)$ to have higher $\epsilon$-order terms.} discussed in Sec. \ref{Sec:nway} and satisfying the rigidity conditions
\begin{equation}\label{Eq:No-rigid_n}
    \int _0^{2\pi}{\rm d}\sigma\ f_i(\sigma) = \int _0^{2\pi}{\rm d}\sigma\ f_i(\sigma) \cos\,\sigma =\int _0^{2\pi}{\rm d}\sigma\ f_i(\sigma) \sin\,\sigma = 0 \ \ \forall \ \ i =1,\cdots n-1.
\end{equation}
These specify unique solutions determined by the gravitational memory $f_i(\sigma)$ in the far past. 

We also obtain the \textit{emergence of $n-1$ correlated clocks $\tau_{d_i}$ at each point in the junction without the need of external observers.} In fact, if
\begin{equation}
    \tau_{d_i}^{(0)}(\sigma) = \lim_{\tau\rightarrow -\infty} \tau_{d_i}(\tau,\sigma),
\end{equation}
then
\begin{equation}
    \tau_{d_i}^{(0)}(\sigma) = f_i'(\sigma) +\mathcal{O}(\epsilon^4)
\end{equation}
implying that the starting values of the clocks are determined by $f_i(\sigma)$. We note that the worldsheet gauge choice \eqref{Eq:WSgauge} and the clocks $\tau_{d_i}(\tau,\sigma)$ at each point on the junction are invariant under physical (proper) spacetime diffeomorphisms generated by vector fields $\xi_i^\mu$ on the $n$-manifolds for which $\xi_i^t$ and $\xi_i^\phi$ vanish at the location of the junction for all $i =1,\cdots,n$. The rigidity conditions \eqref{Eq:No-rigid_n} together with the initial conditions \eqref{Eq:ic-full_n} ensure that these clocks cannot be undone by worldsheet and spacetime isometry transformations, which produce solutions related by the isometries of the trivial maximally symmetric solution in the tensionless limit. 

As mentioned before in Sec. \ref{Sec:nway}, the Nambu-Goto-Monge-Amp\`{e}re equations that appear at the second order in the $\epsilon$ expansion are already non-linear. So we cannot readily verify whether such solutions are well-behaved by obtaining generic exact solutions of these non-linear equations which we have not been able to solve beyond special cases mentioned in Sec. \ref{Sec:nway}. Nevertheless, we have verified via early-time expansion of the solutions of the junction conditions obtained at second and higher orders in $\epsilon$ that multi-valued terms do not appear for the initial conditions given by \eqref{Eq:ic-full_n}. Therefore, the solutions of the $n$-way gravitational junction conditions corresponding to transient vibrations of the coincident $n-1$ string vibrations about the equator are well-behaved. Furthermore, as in the context of the radial expansion AdS \cite{Chakraborty:2025jtj}, the early time expansion reveals that infinite number of non-trivial well-behaved solutions exist even in the tensionless limit (more below).

When $\lambda\neq 0$, the relationship between $\theta_{d,i}^{(0)} = f(\sigma)$ and  
\[
\theta_{d_i}^{(3)}(\sigma) =\lim_{\tau\rightarrow-\infty}\theta_{d_i}(\tau,\sigma) e^{-3\tau}
\]
the leading terms of the stringy vibrations at early time generalize that of the two-way case given by \eqref{Eq:key-relation}. Specifically, we obtain
\begin{align}\label{Eq:key-relation-2}
    \theta_{d_i}^{(3)}
&=
-\frac{8n}{3\lambda}
\Bigg[
\,\sigma_{d_i}^{(0)'}+
\,\sigma_{d_i}^{(0)'''} 
\Bigg]+\mathcal{O}(\epsilon^3) \notag \\&= -\frac{8n}{3\lambda}
\Bigg[
\,f'+
\, f''' 
\Bigg]+\mathcal{O}(\epsilon^3)\ \ \forall \ \ i =1,\cdots n-1,
\end{align}
when $\lambda \neq 0$. This reduces to \eqref{Eq:key-relation} for $n=2$ after replacing $\theta_{d}$ with $-\theta_s$. Recalling that our setup in Sec. \ref{Sec:nway} in which we glued the northern hemispheres of the $n$-manifolds to obtain the $n$-way junction, it is necessary to exchange $\theta_s$ with $-\theta_{d}$ to compare with our results for the 2-way junction where we glued the northern hemisphere of one manifold with the southern hemisphere of the other. 

Crucially, for $\lambda = 0$, we obtain that the leading behavior of $\theta_{d_i}$ at early time is $e^{2\tau}$ instead of $e^{3\tau}$. If
\[
\theta_{d_i}^{(2)}(\sigma) =\lim_{\tau\rightarrow-\infty}\theta_{d_i}(\tau,\sigma) e^{-2\tau}
\]
then we obtain that 
\begin{align}\label{Eq:key-relation-3}
    {\theta_{d_i}^{(2)}}^{2} - \theta_{d_i}^{(2)} \sum_{j \neq i}\theta_{d_j}^{(2)}&= -4 (\sigma_{d_i}^{(0)'}+\sigma_{d_i}^{(0)'''})+\mathcal{O}(\epsilon^3)\notag\\
    &=-4 (f'+f''')+\mathcal{O}(\epsilon^3)\ \ \forall \ \ i =1,\cdots n-1
\end{align}
when $\lambda =0$. In this case, we explicitly obtain how the gravitational memory in the far past, encoded in $f_i(\sigma)$, the relative shifts of the angular coordinates at the junction, encode the transient string vibrations in the tensionless limit. The analogs of the relations \eqref{Eq:key-relation-2} and \eqref{Eq:key-relation-3} relating the displacement operators of the dual interface with the relative time-shifts across the interface in the context of the gravitational junction in AdS$_3$ in the tensionfull and tensionless cases, respectively, can be found in \cite{Chakraborty:2025jtj}.

\section{Discussion and outlook}
\label{Sec:conclusions}

In this work, we have shown that transient vibrations of the strings about the equator in dS$_3$ emerge from well-behaved solutions of multi-way gravitational junctions in dS$_3$. Furthermore, we have demonstrated that these transient string vibrations dynamically set up clocks at the junction in a self-consistent way. These clocks are invariant under proper spacetime diffeomorphisms that vanish at the junction for an appropriate choice of worldsheet gauge. For junctions gluing three or more locally dS$_3$ manifolds, such clocks emerge even in the tensionless limit, establishing a novel feature in pure three-dimensional gravity.

It is clearly important to understand how to use the clocks emerging from the stringy degrees of freedom of the gravitational junction to enable the construction of quantum reference frames \cite{Page:1983uc,Giacomini:2017zju,Hoehn:2019fsy,delaHamette:2020dyi}, and subsequently, gravitationally dressed diffeomorphism invariant approximately local observables \cite{Giddings:2005id,Tambornino:2011vg,Hartle:2015vfa,Chataignier:2020fys,Chandrasekaran:2022cip} in semi-classical quantum gravity in dS space. The key new ingredient provided by the gravitational junction is that \textit{such quantum reference frames arise dynamically in a self-consistent way via the stringy degrees of freedom of the gravitational junction without the need for external observers.} Furthermore, \textit{even in pure gravity, such quantum reference frames can arise dynamically from tensionless junctions gluing three or more locally dS$_3$ manifolds without external observers}. It would be of interest to understand how such constructions generalize to cosmological spacetimes as well. It is also pertinent to mention that the dynamical emergence of quantum reference frames should be crucial for construction of non-perturbative diffeomorphism invariant observables.

It would be of interest to understand whether the gravitational junctions can be interpreted in the language of the dS/CFT correspondence, in line with the decoding of gravitational junctions in AdS in terms of quantum maps characterizing interfaces in the dual CFT \cite{Bachas:2020yxv,Chakraborty:2025dmc,Banerjee:2025zuw,Chakraborty:2026wip}. Also, it would be of interest to examine whether relational observables can be relevant for bulk reconstruction (see \cite{harlow2018tasi,Jahn:2021uqr,Kibe:2021gtw} for reviews) in the context of the dS/CFT correspondence and the generalization of the notion of entanglement wedges. The study of pseudo-entropy in dS$_3$ with gravitational junctions via the holographic prescriptions of \cite{Nakata:2020luh,Doi:2022iyj,Narayan:2022afv,Nanda:2025tid} using methodology developed in \cite{Kibe:2021qjy,Banerjee:2022dgv,Kibe:2024icu} in the context of junctions in AdS$_3$ space can provide some relevant insights. 

Finally, we mention that it would be of interest to study multi-way gravitational junctions in dS$_4$ with the simplical constructions mentioned in \cite{Chakraborty:2025jtj} that allow for the dynamical emergence of stringy degrees of freedom, and therefore quantum reference frames.

\begin{acknowledgments}
AC, AM and MM acknowledge support from FONDECYT postdoctoral grant no. 3230222, FONDECYT regular grant no. 1240955 and ``Doctorado Nacional'' grant no. 21250596 of La Agencia Nacional de Investigaci\'{o}n y Desarrollo (ANID), Chile, respectively. AC appreciates the warm hospitality extended by AM and Instituto de F\'{\i}sica, Pontificia Universidad Cat\'{o}lica de Valpara\'{\i}so, Chile where majority of the work was carried out. AM gratefully acknowledges the hospitality of LPENS, where a substantial part of this work was carried out during his tenure as a CNRS invited professor.
\end{acknowledgments}

\bibliographystyle{JHEP}
\bibliography{ref_PRL}

\end{document}